\documentclass[10pt,conference]{IEEEtran} 
\IEEEoverridecommandlockouts
\usepackage{cite}
\usepackage{amsmath,amssymb,amsfonts}
\usepackage{algorithmic}
\usepackage{graphicx}
\usepackage{textcomp}
\usepackage{xcolor}
\usepackage{orcidlink}
\usepackage[nolist]{acronym}
\usepackage{graphicx}
\usepackage{xcolor}
\usepackage{colortbl}
\usepackage{tikz}
\usetikzlibrary{calc}
\usepackage{standalone}
\usepackage{booktabs}
\usepackage{textpos}
\definecolor{uni-bright}{RGB}{0,190,255}
\definecolor{uni-dark}{RGB}{0,81,158}

\def\BibTeX{{\rm B\kern-.05em{\sc i\kern-.025em b}\kern-.08em
    T\kern-.1667em\lower.7ex\hbox{E}\kern-.125emX}}
\graphicspath{{fig},{tikzfig}}
\begin{document}
\def\ie{i.e.}
\def\N{\mathbb N}
\def\R{\mathbb R}
\def\TM{\texttrademark\ }
\def\1{\mathbf{1}}
\def\0{\mathbf{0}}
\def\I{\mathbb{I}}
\def\mat#1{\ve{#1}}
\def\norm#1{\ensuremath{\left\lVert#1\right\rVert}}
\def\ve#1{\mathlette{\boldmath}{#1}}
\def\pauli#1#2{\ensuremath{\hat{\sigma}_\mathrm{#1}^{(#2)}}}
\def\mathlette#1#2{{\mathchoice{\mbox{#1$\displaystyle #2$}}%
            {\mbox{#1$\textstyle #2$}}%
            {\mbox{#1$\scriptstyle #2$}}%
            {\mbox{#1$\scriptscriptstyle #2$}}}}
\def\diag{\mathrm{diag}}
\def\x{\ve{x}}
\def\Qxx{\ve{Q}_{xx}}
\def\Qxs{\ve{Q}_{xs}}
\def\Qsx{\ve{Q}_{sx}}
\def\Qss{\ve{Q}_{ss}}
\def\Zx{\ve{Z}_{x}}
\def\Zs{\ve{Z}_{s}}
\def\tril{\mathrm{tril}}
\def\triu{\mathrm{triu}}
\def\todo#1{\textcolor{red}{#1}}
\def\colorlabelcircle#1{\tikz{\fill[#1] (0,0) circle (2.5pt)}}

\title{Tactile Network Resource Allocation enabled by Quantum Annealing based on ILP Modeling

    \thanks{The authors gratefully acknowledge the Jülich Supercomputing Centre for funding this project by providing computing time through the Jülich UNified Infrastructure for Quantum computing (JUNIQ) on the D-Wave Advantage\texttrademark\ quantum system.

        This work has been performed in the framework of the CELTIC-NEXT EUREKA project AI-NET ANTILLAS (Project ID C2019/3-3), and it is partly funded by the German Federal Ministry of Education and Research (Project ID 16 KIS 1312).

        The authors thank C. C. Chang and D. Willsch for fruitful discussions.

        The authors alone are responsible for the content of the paper.}
}

\author{
    \IEEEauthorblockN{Arthur Witt\IEEEauthorrefmark{1} \orcidlink{0000-0003-1180-1172},
        Christopher Körber\IEEEauthorrefmark{2} \orcidlink{0000-0002-9271-8022},
        Andreas Kirstädter\IEEEauthorrefmark{3} and
        Thomas Luu\IEEEauthorrefmark{4} \orcidlink{0000-0002-1119-8978}}

    \IEEEauthorblockA{\IEEEauthorrefmark{1},\IEEEauthorrefmark{3}
        \textit{Institute of Communication Networks and Computer Engineering,
            University of Stuttgart}, Stuttgart, Germany}
    \IEEEauthorblockA{\IEEEauthorrefmark{2}
        \textit{Institute of Theoretical Physics II, Ruhr-University Bochum},
        Bochum, Germany}
    \IEEEauthorblockA{\IEEEauthorrefmark{2}
        \textit{Fraunhofer Research Institution for Energy Infrastructures and Geothermal Systems IEG}, Bochum, Germany}
    \IEEEauthorblockA{\IEEEauthorrefmark{4}
        \textit{Institute for Advanced Simulation (IAS-4) \& JARA HPC, Forschungszentrum Jülich},
        Jülich, Germany}
    \IEEEauthorblockA{\{arthur.witt, andreas.kirstaedter\}@ikr.uni-stuttgart.de, christopher.koerber@ieg.fraunhofer.de, t.luu@fz-juelich.de}
}

\maketitle
\begin{textblock}{13.3}(0,-4.25)
    \textcolor{black}{This work has been submitted to the IEEE for possible publication. Copyright
        may be transferred without notice, after which this version may no longer be
        accessible.}
\end{textblock}
\begin{acronym}[WWW]
    \acro{DWDM}{dense wavelength division multiplexing}
    \acro{ILP}{integer linear program}
    \acro{IP}{internet protocol}
    \acro{MDS}{minimum dominating set}
    \acro{NP}{non-polynomial}
    \acro{OXC}{optical cross connects}
    \acro{QA}{quantum annealer}
    \acro{QC}{quantum computing}
    \acro{QUBO}{quadratic unconstrained binary optimization}
    \acro{QPU}{quantum processing unit}
    \acro{SDN}{software-defined networking}
    \acro{uRLLC}{ultra reliable and low latency communication}
    \acro{WDM}{wavelength division multiplexing}
\end{acronym}

\vspace{-4mm}

\begin{abstract}
    Agile networks with fast adaptation and reconfiguration capabilities are required for on-demand provisioning of various network services.

    We propose a new methodical framework for short-time network optimization based on \acf{QC} and \acf{ILP} models, which has the potential of realizing a real-time network automation. We define methods to map a nearly real-world \ac{ILP} model for resource provisioning to a \acf{QUBO} problem, which is solvable on \acf{QA}.

    We concentrate on the three-node network to evaluate our approach and its obtainable quality of solution using the state-of-the-art quantum annealer D-Wave Advantage\TM 5.2/5.3.  By studying the annealing process, we find annealing configuration parameters that obtain feasible solutions close to the reference solution generated by the classical \ac{ILP}-solver CPLEX.

    Further, we studied the scaling of the network problem and provide estimations on quantum annealer's hardware requirements to enable a proper \ac{QUBO} problem embedding of larger networks. We achieved the \ac{QUBO} embedding of networks with up to 6 nodes on the D-Wave Advantage\texttrademark. According to our estimates a real-sized network with 12 to 16 nodes require a \ac{QA}  hardware with at least 50000 qubits or more.
\end{abstract}


\section{Introduction}

\subsection{Motivation}
Optical wide-area networks are the backbone for public communication systems like 5G/6G mobile communication and different variants of fixed-access networks.
The transport of \ac{IP} traffic requires a conversion from electrical to optical signals and vice versa.
It is realized by power-hungry transceivers within the optical networks.
Because traffic volume changes over time, adapting the network configuration to new demands is beneficial.

Traffic temporal changes occur regularly on a diurnal scale \cite{bib:Traffic_IKR,bib:Traffic_Automotive_Application} but also have fluctuations at smaller timescales of seconds and less \cite{bib:Traffic_Stanford, bib:Traffic_Long_Term_Study}.
Therefore, an economically-friendly adaptive operation of networks requires fast control algorithms for dynamic resource allocation, traffic engineering and restoration.

The underlying problem of network resource adaptation is a constrained optimization problem of integer variables.
This kind of problem can be formulated as an \acf{ILP}.
Solving \acp{ILP} for network resource allocation can be time consuming, especially if a detailed modeling of the network architecture and operation strategies is required, which is usually the case.
Therefore, an \ac{ILP}-based network adaptation in short time periods is difficult to achieve---in \cite{bib:ILP_Tornatore} solutions are obtained within several hours.
In a real network, problems like these are often solved with heuristic or meta-heuristic approaches which allow for a fast solvability but at the expense of modeling accuracy.
These methods also require substantial effort to tailor the design of the heuristic to the specific problem at hand. Those comparatively fast heuristics require still several minutes to solve such problems \cite{bib:Heuristic_Feller,bib:Heuristic_Bauknecht}.

Quantum hardware that leverages the superposition of quantum bits enables massively parallel optimization protocols and could possibly overcome the time constraints for short time optimizations without the need for reduced model accuracy.  
For example, refs.~\cite{bib:McGeoch17} and \cite{bib:quantum_supremacy} showed that classical solution approaches can be outperformed by quantum computing.  
A significant optimizer speedup on the order of magnitudes is, in principle, achievable.
Only through recent technological progress has it become possible to map non-trivial optimization problems on quantum hardware.
In particular, our work proposes a new \ac{ILP}-based solution approach for the aforementioned network problem that can be solved on a \acf{QA}, a form of quantum computing architecture.
While solutions obtained on a quantum hardware are not generally guaranteed to be optimal, this problem is an ideal use case for such algorithms as it is only required to obtain better solutions---which can be verified within microseconds.

\subsection{Objectives}
We test the feasibility of solving network optimization problems on quantum hardware.
Specifically, we test the feasibility to extract solutions of a small but non-trivial discrete optimization problem within seconds.
Because today's networks are controlled centrally via \acf{SDN}, it is technically possible to incorporate this quantum framework in realistic network configuration setups for traffic engineering and restoration.

We use the new D-Wave Advantage\TM 5.2/5.3 quantum annealer in Jülich for the evaluation of our network optimizing algorithm.
We illustrate the essential steps in the general application of the \ac{QA} to the network problem, and discuss the findings, discovered challenges, and restrictions of our approach for a three-node network.
Additionally we provide a network scaling study which allows us to make quantitative statements related to the feasibility of our approach and \ac{QA}-related requirements for solving real-world network problems.

\subsection{Organization of the Work}
Our work covers the interdisciplinary aspects of quantum-aided optimization and network resource allocation as problem. The allocation problem is introduced as a nearly real-world application of quantum computing. 
In particular, we specify the considered network architecture in Sec.~\ref{sec:formulation-rap} and formulate the resource allocation problem of wide-area networks.
In Sec.~\ref{sec:solution_approach}, we introduce our proposed algorithmic methods to solve a typical network resource allocation problem with quantum annealing.
Therefore, we model the resource allocation problem as an \ac{ILP} and explicitly derive the formulation of this ILP in terms of a \ac{QUBO} representation required for the optimization procedure on the \ac{QA}.
In Sec.~\ref{sec:Evaluation} we  evaluate our solution method according to a defined procedure and interpret the results that we obtained for a three-node network.
We also discuss the stochastic effects of the quantum annealing process.
In Sec.~\ref{sec:scalability-study} we introduce the embedding of QUBO matrices on the D-Wave \ac{QA} hardware and study the scalability of our approach with respect to increased network sizes and hardware limitations.
This allows us to estimate the QA hardware requirements for real-sized networks.
Finally, we recapitulate in Sec.~\ref{sec:conclusion}.
Additionally, we provide supplemental material in the appendices on the idea of quantum annealing (Appx.~\ref{sec:app_dwave}) and a summary on the general approach of ILP to QUBO mapping (Appx.~\ref{sec:app_problem_mapping}).

\section{Resource Allocation in Wide Area Networks as Combinatorial Optimization Problem}
\label{sec:formulation-rap}

\subsection{Network Architecture}
An optical wide-area network is structured like a graph with a meshed topology consisting of network nodes that are linked by optical fiber systems on the graph's edges. 
An optical fiber system itself, denoted in short as a fiber link, is a sequence of fiber spans (each with 80km length) with amplifiers in between. 
Edges within a wide-area network allow transmissions in both directions. 
They are typically realized by separated fibers to avoid signal distortions due to crosstalk and other effects \cite{bib:bidirect_transmission}.

Individual signals, propagating in the same direction, can be transported simultaneously on a fiber using \acf{DWDM}. 
The fiber bandwidth is divided into a finite number of channels each with $50$GHz bandwidth. 
An optical channel is specified by its central wavelength. 

Network nodes $V$ are composed of electrical \ac{IP} routers in an upper layer and optical equipment, like \ac{OXC}, in a lower layer. 
Transition between these layers, i.e., the electrical-to-optical signal transformation and the other way round, are enabled by power-hungry transceivers. 
They allow transmission (TX) and reception (RX) via time-division multiplexing, 
but are used mostly unidirectionally. 
An optical signal generated by a transceiver utilizes a spectral bandwidth dictated by the granularity of that single optical channel.
Within the \ac{OXC}, optical signals between transceivers and connected fiber links are switched in a wavelength-selective way.
In particular, redirected signals can be forwarded only at the same wavelength.
Signals at the fiber link side are de-/multiplexed to apply \ac{DWDM} signals on fiber links.

\begin{figure}
    \includegraphics[width=\columnwidth]{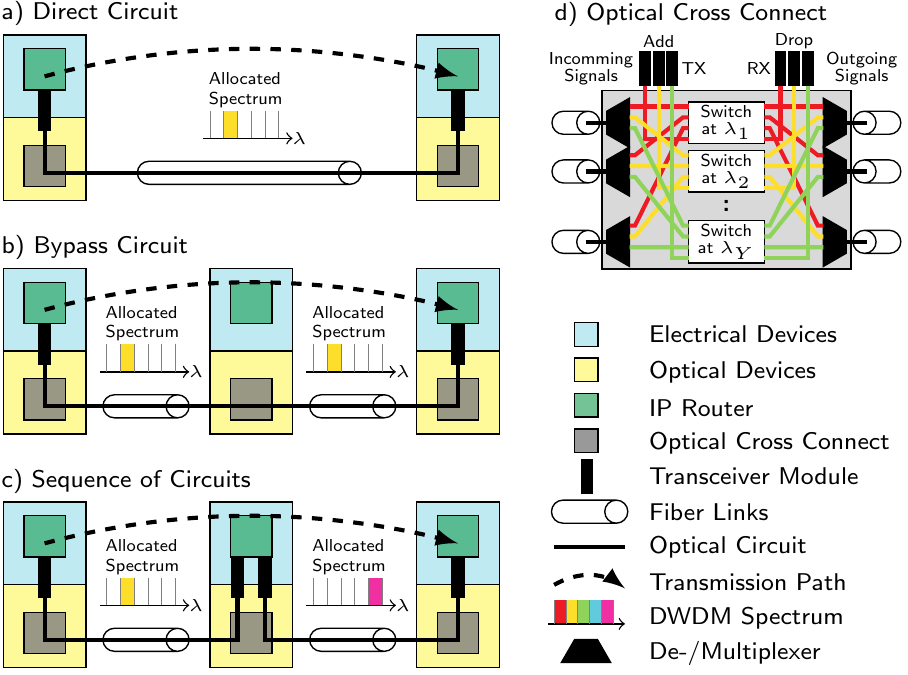}
    \caption{a--c) Various ways of realizing transmission paths in wide-area networks with optical \acf{DWDM} layer and d) architecture of a \acf{OXC}.}
    \label{fig:tp_realization}
\end{figure}

The transportation of data requires the realization of a transmission path. 
A transmission path is a list of traversed nodes which includes the source and target node. 
It is realized by a single circuit or a sequence of optical circuits. 
Optical circuits are terminated by transceivers and utilize the spectrum of a single optical channel. 
A \emph{direct} optical circuit traverses a single fiber link whereas a \emph{bypass} circuit traverses at least two fiber links (wavelengths are preserved).
If transmission paths are realized by a sequence of optical circuits by a sequence of optical circuits, the allocated wavelength of subsequent circuits may differ.
These variants of transmission path realizations and the functionality of \acp{OXC} are depicted in Fig.~\ref{fig:tp_realization}. 

We will consider a simplified network architecture that is restricted as follows:
\begin{itemize}
    \item \textbf{Singe Rate System:}
    All optical circuits operate at a single data rate of $100\,$Gbit/s with a maximal optical reach of  $1000\,$km. This are typical values in wide-area networks and depend on the type of equipped transceivers. 
    \item \textbf{Colorless Wavelength Assignment:}
    Fibers are typically over-provisioned during the construction phase of a network, such that spectral resources at fiber links are not limited. Therefore, we will ignore the wavelength assignment as it can be optimized in an additional step by a graph coloring algorithm. 
\end{itemize}
Beside these considerations, multiple optical circuits can be realized along the same circuit path $c$ (sequence of traversed nodes). As we ignore the wavelength assignment, we can summarize the specification of those circuits by its unique circuit path $c$ and a circuit counter $w_c$, which will be used to determine the required capacity for a traffic flow or a migration of traffic flows along a circuit path.

\subsection{Connectivity as a Network Service}
Traffic flows from/towards connected data centers and other interconnecting networks ingress and egress the wide-area network at one of the IP routers. A traffic ingress/egress on the optical layer is only possible with additional considerations that are not covered here. Incoming traffic flows are combined by a migration process and considered as single flow if they have to traverse the same source node $u$ and target node $v$ for $u,v\in V$. 
So, we can define a unidirectional traffic demand $d$ for each disjunct node pair ($u,v$) with $u\neq v$. 
The data volume for transport between those nodes---the demand value $h_d$---is expressed as a data rate. 
Demands are summarized in the set $D$ and can be interpreted as a request of connectivity with a specific data rate.

Typically the demands will vary over time, such that a network is adapted by a frequent or on-demand traffic engineering to provide the required capacity and connectivity. 
This process is an essential part of the network operation and considers:
\begin{itemize}
    \item \textbf{Traffic Routing:}
          A transmission path (i.e., sequence of traversed nodes) has to be selected for realizing a traffic flow between two nodes.
          Usually, the shortest path is selected as, in most cases, it provides the shortest latency.
          However, to potentially allocate network resources more efficiently, we consider alternative paths --- which increases the amount of combinations for possible traffic flow migrations.
    \item \textbf{Resource Allocation:}
          Establishing a transmission path requires the activation of optical circuits, see Fig.~\ref{fig:tp_realization}.
          This activation includes the allocation of optical channels per traversed fiber link and transceivers.
          The (expensive) transceivers are node-wise limited and can be used for transmission or reception.
          The considerations for spectrum allocation are already described in the context of the colorless wavelength assignment.
\end{itemize}



\subsection{Definition of Possible Transmission Path Realizations}
\label{sec:Paths}

Initially, we generate a set of circuit paths $C$ that includes all possible direct circuit paths that traverse a single network edge and a finite number of possible alternatives in the form of bypass circuit paths.
We must therefore consider the optical reach of a circuit and the restrictions of the topology.

Based on a shortest path search, we can identify a set of one or more loop-free transmission paths $L_d$ per demand $d$, that are listed by their distance in ascending order.
It can be beneficial if some of the transmission paths are realized by a sequence of rather short circuit paths.
If this is the case, it increases the flexibility to carry a combination of different traffic flows along a circuit path $c$ by data migration.
Therefore, we compute the set $R_{d,l}$ of possible circuit path patterns $r\in R_{d,l}$, that realize transmission path $l\in L_d$. 
The pattern $r$ consists of a sequence of circuit paths or a single circuit path $c$, predefined in $C$.
The first circuit path pattern in $R_{d,l}$ is composed of the shortest possible circuit paths, i.e. they traverse only a single edge of the network.
The second pattern uses a minimal amount of circuit paths which incorporate multiple edges and therefore increased distances.  
Further patterns have mixed circuit path lengths, and so on.
For simplicity we will define the union set $T_d$, where $t_d\in T_d$ are all circuit path patterns that can realize a transmission path for demand $d$, and the total union set $T$ according to
\begin{equation}
    T_d=\bigcup\limits_{l\in L_d}R_{d,l} \quad \text{and}\quad
    T=\bigcup\limits_{d\in D}T_d.
\end{equation}
Circuit paths of $C$, that are not used in $T$ are deleted in $C$.

The economic resource allocation within \ac{WDM} networks, in this context, forms a combinatorial optimization problem with a \acf{NP}-complete complexity.

\section{Algorithmic Solution Approach}
\label{sec:solution_approach}

\subsection{Framework}
\label{sec:Framework}
\begin{figure}
    \includegraphics[width=\columnwidth]{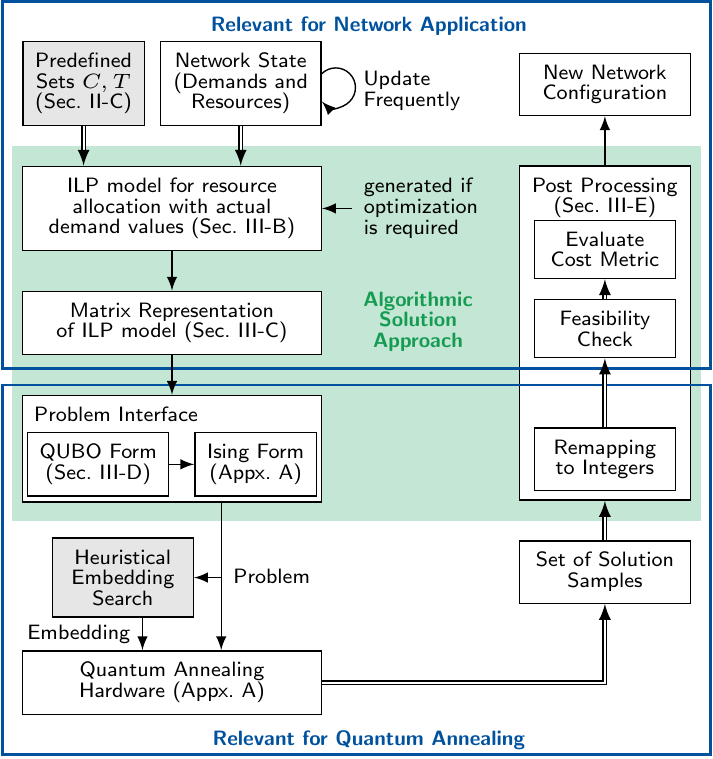}
    \caption{Proposed framework for network resource allocation enabled by ILP modeling and quantum annealing at a glance. Gray boxes represent work steps which need to be executed once in a setup phase. White boxes form the control loop of regularly optimization, which is triggered frequently or on-demand.}
    \label{fig:framework}
\end{figure}

Optimizing the resource allocation problem of wide-area networks involves multiple interacting stages: a) the modeling of the network as an \ac{ILP} problem, b) finding a representation of this problem in \ac{QA}-compatible fashion, c) finding possible embeddings of the problem on the \ac{QA}, d) evaluating the output of the \ac{QA}, and e) performing a post-processing analysis to find feasible solutions that optimize the network.
Some of these steps are time intensive but only need to be executed once, other steps need to be performed multiple times but can be executed within microseconds.
Fig.~\ref{fig:framework} summarizes the concept of our approach for solving network resource allocation problems by quantum annealing.

Our solution approach relies on a method for solving \acfp{ILP} on quantum annealers that was published firstly in \cite{bib:QAasILP}.
The authors propose a strategy for mapping an \acf{ILP} to a \acf{QUBO} problem and used it to solve a binary linear problem.
In this work, we incorporate a mostly real-world network resource allocation problem.
This allows us to study this mapping process and estimate the feasibility of incorporating this formalism for a real world use case.
We also give an extended form of the generic matrix-wise definition of the \ac{QUBO} representing non-binary \ac{ILP} target variables.

\subsection{Integer Linear Program for Network Optimization}
\label{sec:ILP}
\Acfp{ILP} for resource allocation and service provisioning in wide area networks are often used as a reference method providing exact solutions for comparison with newly developed heuristic or meta-heuristic algorithms. Sometimes they are also used to study the possible benefit of new network operation strategies. Some examples are given in \cite{bib:IKR1,bib:IKR2} where the reduction of over-provisioned quality of services in networks was studied.

To study the applicability of solving a network resource allocation problem on quantum annealers we define the following terms based on \cite{bib:IKR2}.
Though the traffic volume $h_d$ of a demand $d$ varies over time, for the \ac{ILP} it is held constant. 
The variations over time can be considered by frequently solving the \ac{ILP} with updated values of $h_d$.

We now enumerate the parameters and constraints of our ILP.

\vspace{2mm}\noindent\textit{Variables:}
\begin{itemize}
    \setlength\itemsep{-0.2em}
    \item $g_{t_d} \in \{0,1\}$: path selector, \ie, $g_{t_d}$ equals 1 if a transmission path for demand $d$ is realized by circuit configuration $t_d\in T_d$.
    \item $w_c \in \N$: the number of active, parallel circuits on circuit path $c$.
\end{itemize}
\textit{Constants:}
\begin{itemize}
    \setlength\itemsep{-0.2em}
    \item $\xi \in \R$: the data rate of a single optical circuit.
    \item $\eta_v \in \N$: the amount of transceivers installed at node $v$.
    \item $\rho_{c,t_d} \in \{0,1\}$: indicates whether circuit configuration $t_d$ uses circuit path $c$.
    \item $\varphi_{v,c} \in \{0,1\}$: indicates whether node $v$ is the source or target node of circuit path $c$.
    \item $h_d\in \mathbb{R}$: traffic volume of demand $d$.
\end{itemize}
\textit{Constraints}:
\newlength{\negspace}
\setlength\negspace{0mm}
\begin{eqnarray}
    \hspace{-5mm}\sum_{t_d \in T_d} g_{t_d} = 1 \hspace{\negspace}&&\forall d \in D\hspace{5mm}
    \label{eq:const_demand}\\
    -w_c +
    \sum_{d \in D}
    \sum_{t_d \in T_d}
    \rho_{c,t_d} \cdot\frac{h_d}{\xi} \cdot g_{t_d} \le 0 \hspace{\negspace}&&\forall c \in C
    \label{eq:const_circuits}\\
    \sum_{c \in C} w_c \cdot   \varphi_{v,c} \le \eta_v  \hspace{\negspace}&&\forall v \in V
    \label{eq:const_nodes}
\end{eqnarray}
\textit{Objective:}
\begin{eqnarray}
    \sum_{c \in C} w_c\ \rightarrow\ \min.\label{eq:objective}
\end{eqnarray}
Equation~\eqref{eq:const_demand} ensures that a demand is routed on exactly one path. The constraint~\eqref{eq:const_circuits} ensures that enough circuits are activated depending on the chosen paths.
Constraint~\eqref{eq:const_nodes} activates a sufficient amount of transceivers to accommodate the active optical circuits. Finally, the objective~\eqref{eq:objective} minimizes the number of active circuits and therefore also the number of active transceivers.

\subsection{Network-related ILP Model in Matrix Form}
\label{sec:network_matrix}

An intermediate step of the \ac{ILP} to \ac{QUBO} mapping described in \cite{bib:QAasILP} is the reformulation of the \ac{ILP} in matrix form.
In general, an \ac{ILP} can be defined by an objective function that incorporates a set of $K$ constraints that are defined with $M$ integer variables $x_m \geq 0$.   We collectively describe these integer variables in vector form, $\ve{x}\in \mathbb{N}^M$. The objective function defines our optimization target, which in this case involves a minimization,
\begin{equation}
    \ve{c}^\top \ve{x}\rightarrow \min\ ,
    \label{eq:cost_function}
\end{equation}
where $\ve{c}^\top \in \mathbb{R}^M$ represents our cost weights. The constraints are given in the form of equalities or inequalities according to
\begin{equation}
    \label{eq:ilp-constraint}
    \ve A \ve{x} + \ve b \leq 0
\end{equation}
with constant values $\ve{b}\in\mathbb{R}^K$ and the matrix $\mat{A}$ with shape $K\times M$ that contains the linear weights $A_{k,m}\in \mathbb{R}$ for all constraints of the \ac{ILP}.
Typically, when solving \acp{ILP}, the inequalities in~\eqref{eq:ilp-constraint} are transformed into equalities by introducing integer slack variables $\ve{s}\in \mathbb{N}^K$ such that
\begin{equation}
    \begin{aligned}
        \ve A \ve{x} + \ve b & \leq 0
        \Leftrightarrow
        \mat A \ve{x}+\ve{b}+\ve{s}=\ve{0}\ .
    \end{aligned}
\end{equation}
Thus, the network allocation \ac{ILP} given in Sec,~\ref{sec:ILP} is equivalent to
\begin{equation}
    \ve A=
    \begin{bmatrix}
        \ve{G}^{|D|\times|T|} &
        \0^{|D|\times|C|}           \\
        \ve{H}^{|C|\times|T|} &
        -\I^{|C|\times|C|}          \\
        \0^{|V|\times|T|}     &
        \ve{\varphi}^{|V|\times|C|} \\
    \end{bmatrix},
    \ve{b}=-\begin{bmatrix}
        \1^{|D|}        \\
        \0^{|C|}        \\
        \ve{\eta}^{|V|} \\
    \end{bmatrix},
    \ve{s}=\begin{bmatrix}
        \0^{|D|}                   \\
        {\ve{s}_\mathrm{c}}^{|C|}  \\
        \ve{s}_\mathrm{v}^{|V|} \\
    \end{bmatrix},\nonumber
\end{equation}
\begin{equation}
    \ve{x}=\begin{bmatrix}
        \ve{g}^{|T|}      \\
        \ve{\omega}^{|C|} \\
    \end{bmatrix},\hspace{1mm}
    \ve{c}^\top=\begin{bmatrix}
        \0^{|T|} &
        \1^{|C|}
    \end{bmatrix}.
\end{equation}

The term $|\cdot|$ in the superscript of the matrices refers to the size of the used sets $C,D,T$ and $V$ and thus defines the matrices' dimensions. 
The expression $\0$ refers to a vector or matrix that contains only zero values, while the entries of $\1$ are all one. 
$\I$ is the identity matrix. 
Note that the column vector $\ve{x}$ contains the \ac{ILP}'s variables $\ve{g}=[g_{t}]_{t \in T},\ g_t\in\{0,1\}  $ and $\ve{\omega}=[\omega_c]_{c\in C},\ \omega_c\in\{1,2,\ldots, \omega_\mathrm{max}\}$. 

The rows of $\ve{A}$, $\ve{b}$ and $\ve{s}$ are ordered according to the constraints of the \ac{ILP}. The first row with matrix
\begin{equation}
    \ve{G}=\begin{bmatrix}
        \1^{1\times|T_1|} &
        \0^{1\times|T_2|} &
        \ldots            &
        \0^{1\times|T_{|D|}|} \\\
        \0^{1\times|T_1|} &
        \1^{1\times|T_2|} &
        \ldots            &
        \0^{1\times|T_{|D|}|} \\
        \vdots            &
        \vdots            &
        \ddots            &
        \vdots                \\
        \0^{1\times|T_1|} &
        \0^{1\times|T_2|} &
        \ldots            &
        \1^{1\times|T_{|D|}|} \\
    \end{bmatrix}
\end{equation}
describes the possible path selection according to \eqref{eq:const_demand}. The second row describes how the traffic volume $h_d$ 
is distributed over the set of circuits $C$, cf.~\eqref{eq:const_circuits}.
As $h_d\in\mathbb{R}$ does not necessarily take on discrete values, we introduce a discretized version $\bar h_d$ for our \ac{ILP} by designating a desired number of representative digits $a\in\mathbb{N}$ in conjunction with the normalization to the single circuit capacity $\xi$, 
\begin{equation}
    \overline{h}_d=\left\lceil\frac{h_d*2^a}{\xi}\right\rceil \cdot \frac{1}{2^a}.
\end{equation}
Here the expression $\lceil\cdot\rceil$ indicates a rounding to the next largest integer value. 
Together with the binary matrix $\ve{\rho}_d=[\rho_{c,t_d}]_{c_\in C, t_d \in Td}$ that describes the existence of a circuit $c$ inside a circuit configuration $t_d$ for demand $d$, the matrix H is given as
\begin{equation}
    \ve{H}=\begin{bmatrix}
        \overline{h}_1\ve{\rho}_{1} & \overline{h}_2\ve{\rho}_{2} & \ldots & \overline{h}_{|D|}\ve{\rho}_{|D|}\
    \end{bmatrix}\ .
\end{equation}
The last rows of $\ve{A}$, $\ve{b}$ and $\ve{s}$ incorporate the limited amount of installed transceivers $\ve{\eta}=[\eta_v]_{v\in V},\ \eta_v\in \{0,1,\ldots,\eta_\mathrm{max}\}$ per node $v$, cf. \eqref{eq:const_nodes}. 
As such, $\ve{\varphi}=[\varphi_{v,c}]_{v\in V,c\in C}$ describes whether a transceiver at node $v$ is connected to a circuit $c$ binary.

The slack vector $\ve{s}$ has an all-zero block entry in the first block since \eqref{eq:const_demand} is already an equality constraint. 
The remaining blocks $\ve{s}_\mathrm{c}=[s_c]_{c\in C},\ s_c\in\{0,1,\ldots,a\}$ and $\ve{s}_\mathrm{v}=[s_\eta]_{v\in V},\ s_v\in\{0,1,\ldots, \eta_\mathrm{max}\}$ are variables with a limited integer space.

\subsection{Programming the Resource Allocation Problem on the D-Wave Quantum Annealer}
\label{sec:qubo_solving}

The programming of D-Wave’s quantum annealer requires our \ac{ILP} problem to be cast in \acf{QUBO} form
\begin{equation}
    X^2(\ve{q})=\ve{q}^\top \ve{Q} \ve{q} \,
\end{equation}
for bit vectors $\ve{q} \in \{0, 1\}^N$.
In general, given a \ac{QUBO} matrix $\ve Q$, the embedding of the matrix $\ve Q$,  i.e. the mapping of the problem matrix to the annealer qubit topology, is also required.
Such embeddings are generated heuristically by the D-Wave API.
Both $\ve Q$ and its proper embedding are then submitted to the quantum annealer for optimization, which consists of finding the optimal bit vector $\ve{q}$ which minimizes the objective function $X^2$. 
Further details on how the D-Wave quantum annealer solves the \ac{QUBO} problem are given in Appx.~\ref{sec:app_dwave}.

Our network problem can be expressed in \ac{QUBO} form using the matrices $\ve{A}$, $\ve{b}$ and $\ve{s}$ we introduced earlier \cite{bib:QAasILP}.  
We have
\begin{equation}
    X^2(\ve{q})=\ve{q}^\top \ve{Q} \ve{q}+ C
    \label{eq:qubo_energy}
\end{equation}
with
\begin{equation}
    \ve{Q}=p
    \begin{bmatrix}
        \Qxx & \Qxs \\
        \Qsx & \Qss \\
    \end{bmatrix}
    ,\quad C=p \norm{\ve{b}}^2\nonumber
\end{equation}\vspace{-0.5cm}
\begin{eqnarray}
    \Qxx&\hspace{-2mm}=\hspace{-2mm}&\Zx^\top A^\top A \Zx + \diag\left\{\left(2\ve{b}^\top A +\frac{1}{p}\ve{c}^\top\right) \Zx    \right\}
    \nonumber\\
    \Qxs&\hspace{-2mm}=\hspace{-2mm}&\Qsx^\top=\Zx^\top A^\top \Zs
    \nonumber\\
    \Qss&\hspace{-2mm}=\hspace{-2mm}&\Zs^\top \Zs+2\diag\left\{\Zs^\top \ve{b} \right\}\,.
    \label{eq:qubo_matrix}
\end{eqnarray}
The matrices $\mat{Z}_x$ and $\mat{Z}_s$ are defined for mapping between integer and binary variables
\begin{equation}
    \ve{x}=\mat{Z}_x\ve{q}_x\,, \quad \ve{s}=\mat{Z}_s\ve{q}_s\,.
\end{equation}
Since a positive integer variable $x_i$ with an upper limit of $x_{i,\mathrm{max}}$ can be expressed by $R_i=\left\lceil\log_2(x_{i,\mathrm{max}})\right\rceil$ binary values, we can express $\ve z_i$ as
\begin{equation}
    \ve z_i=\begin{bmatrix}
        2^{R_i-1} & 2^{R_i-2} & \ldots & 2^{1} & 2^0
    \end{bmatrix}\,.
    \label{eq:z_indv}
\end{equation}
The mapping matrix for the integer vector $\ve x$ with length $N_x$ can then be defined by
\begin{equation}
    \ve Z_x =
    \begin{bmatrix}
        \ve z_1 & \ve 0   & \ldots & \ve 0       \\
        \ve 0   & \ve z_2 & \ve 0  & \ve 0       \\
        \ve 0   & \ve 0   & \ddots & \ve 0       \\
        \ve 0   & \ldots  & \ve 0  & \ve z_{N_x}
    \end{bmatrix}\,,
    \label{eq:z_comp}
\end{equation} 
and a similar construction can be done for $\ve s$.
The binary variable vectors $\ve q$ themselves can be decomposed as
\begin{equation}
    \ve q=\begin{bmatrix}
        \ve q_x \\
        \ve q_s
    \end{bmatrix}\,.
    \label{eq:q_vector}
\end{equation}
Finally, we have introduced a penalty factor $p$ that weights the relation between the \ac{ILP}'s objective function $\ve c^\top \ve x$ and the error metric $\norm{\ve A\ve x+\ve b+\ve s}$.
In general the penalty factor rates the fulfillment of the \ac{ILP}'s constraints during optimization.
Too small a value of $p$ results in more \ac{QA} solutions that do not satisfy the imposed constraints, whereas too large a value reduces the quality of solutions. 
Our experience has been that the penalty parameter $p$ should be chosen in the range $1\leq p\leq 10$. 

The exact global minimization of~\eqref{eq:qubo_energy} is equivalent to solving our ILP problem.  
A more formal derivation and discussion of this transformation into \ac{QUBO} form is given in  Appx.~\ref{sec:app_problem_mapping}.

\subsection{Post Processing}
The optimization of a single \ac{QUBO} matrix $\ve{Q}$ by quantum annealing returns a set with a large number (hundreds or thousands) of solution samples in a second. The post processing starts with the remapping of binary solution samples to integers. Then, the solutions are rated according to the obtained cost value and their feasibility. Based on the definitions from Sec.~\ref{sec:ILP} and Sec.~\ref{sec:network_matrix}, the solution's feasibility is given if variable vectors $\ve{g}$ and $\ve{\omega}$ fulfill the conditions
\begin{equation}
    \ve{Gg}\overset{!}{=}\1,
    \quad
    \ve{H}\ve{g}-\ve{ \omega}
    \overset{!}{\leq}\0,
    \quad
    \ve{\varphi}\ve{\omega}-\ve{\eta}
    \overset{!}{\leq}\0.
    \label{eq:feasibility_check}
\end{equation}
Finally, a feasible solution sample with the lowest cost value of a solution sample set will be used as the optimized network configuration.

\section{Evaluation of the Solution Approach}
\label{sec:Evaluation}
\subsection{Procedure of Evaluation}
\label{sec:solution-strat}
We evaluate our solution approach described in Sec.~\ref{sec:solution_approach} as follows:
\begin{enumerate}
    \item For a three-node network with normal-distributed demand values $h_d \in \mathcal{N}(75,10)$ in Gbit/s, we generate \acp{QUBO} with different penalty terms $p\in\{1,2,4,8,16,1000\}$ and number of slack accuracy digits $a\in\{1,2,3,4,5\}$. Both parameters can be seen as a knob to tune the numerical accuracy and will have an effect on the \ac{QUBO}'s entries. Parameter $a$ will also affect the shape and structure of the \ac{QUBO}.
    \item \acp{QUBO} with the same penalty, shape and structure are categorized within the same \ac{QUBO} class.
    \item We generate a few different embeddings per \ac{QUBO} class with the heuristic approach \emph{minorminer} provided by the D-Wave's SDK. 
    In principle, an embedding can be used for different \ac{QUBO} matrices of the same class.
    \item We submit the \ac{QUBO} and its proper embedding with varying annealing parameters like chain strength, annealing schedules, and others.
          The annealing schedules summarizing the annealing duration, annealing pause duration and annealing fraction of pausing where the annealing pause is applied. We vary the annealing duration per sample within $t_\mathrm{ps} \in\{1,50,100,500,1000\}$ \textmu s and the annealing fraction of pausing between $s_\mathrm{p}=0.35$ and $0.5$. The pause duration of $t_\mathrm{p}=20$ \textmu s allows a longer interaction duration for the transverse Ising process (see. Appx.~\ref{sec:app_dwave}). 
          We denote the annealing schedule in short by $t_\mathrm{ps} \text{@} s_\mathrm{p} + t_\mathrm{p}$.
    \item For post-processing, obtained binary solutions are translated to integer values. They are classified as feasible if they fulfill the \ac{ILP}'s constraints. We further evaluate the \ac{QUBO}'s energy (value of $\ve{q}^\top \ve{Q} \ve{q}+ C$), the \ac{ILP}'s objective function (value of $\ve c^\top \ve x$) and the number of occurrences per time to solution.
\end{enumerate}

In total, we gathered $5.1$ million solution samples distributed over $3300$ submissions for $40$ different embeddings, resulting in over $2000$ distinct parameter combinations.
We observed that the parameters of the annealing schedule and penalty have a larger impact on solvability. 
Table~\ref{tab:parameters} provides the number of solution samples for a selection of parameter configurations. 
The cells highlighted with a green color are parameter combinations which returned feasible solutions.

\begin{table}
\def\cec{\cellcolor{green!25}}
\def\mus{\,\textmu s}
\caption{Number of 1000 samples per penalty term and annealing schedule. Green entries found feasible solutions.}
\label{tab:parameters}
\centering
\begin{tabular}{l | rrrrrr}
\toprule
Annealing Time or       & \multicolumn{6}{c}{QUBO Penalty $p$}\\
Annealing Schedule      & 1         & 2  & 4        & 8  & 16       & 1000  \\
\midrule
1\mus                   & \cec 600  &    & 170      &    &          & 48    \\
50\mus                  &           &    & 117      &    &          &       \\
100\mus                 &           &    & \cec 99  &    &          &       \\
100\mus @ 0.35 + 20\mus & 50        & 50 & \cec 50  & 50 & 50       &       \\
100\mus @ 0.50 + 20\mus & 50        & 50 & \cec 60  & 50 & \cec 50  &       \\
500\mus                 &           &    & 30       &    &          &       \\
1000\mus                &           &    & 17       &    &          &       \\
\bottomrule
\end{tabular}
\end{table}

The complete set of used run parameters, specifications and obtained solutions are logged to a database and is openly accessible \footnote{\textbf{Open Data Access: }\url{https://jugit.fz-juelich.de/qnet-public/home/}}. The database is realized with EspressoDB \cite{bib:EspressoDB}.

\subsection{Interpretation of the Obtained Solutions}
\begin{figure}
    \centering
    \begin{tikzpicture}[font=\sffamily\fontsize{7pt}{7pt}\selectfont, inner sep=0pt]
        \useasboundingbox (-4.05,-1.93) rectangle (2.3,2.05);
        \node at (0,0)[align=left, inner sep=0pt]{
            \includegraphics[width=\columnwidth]{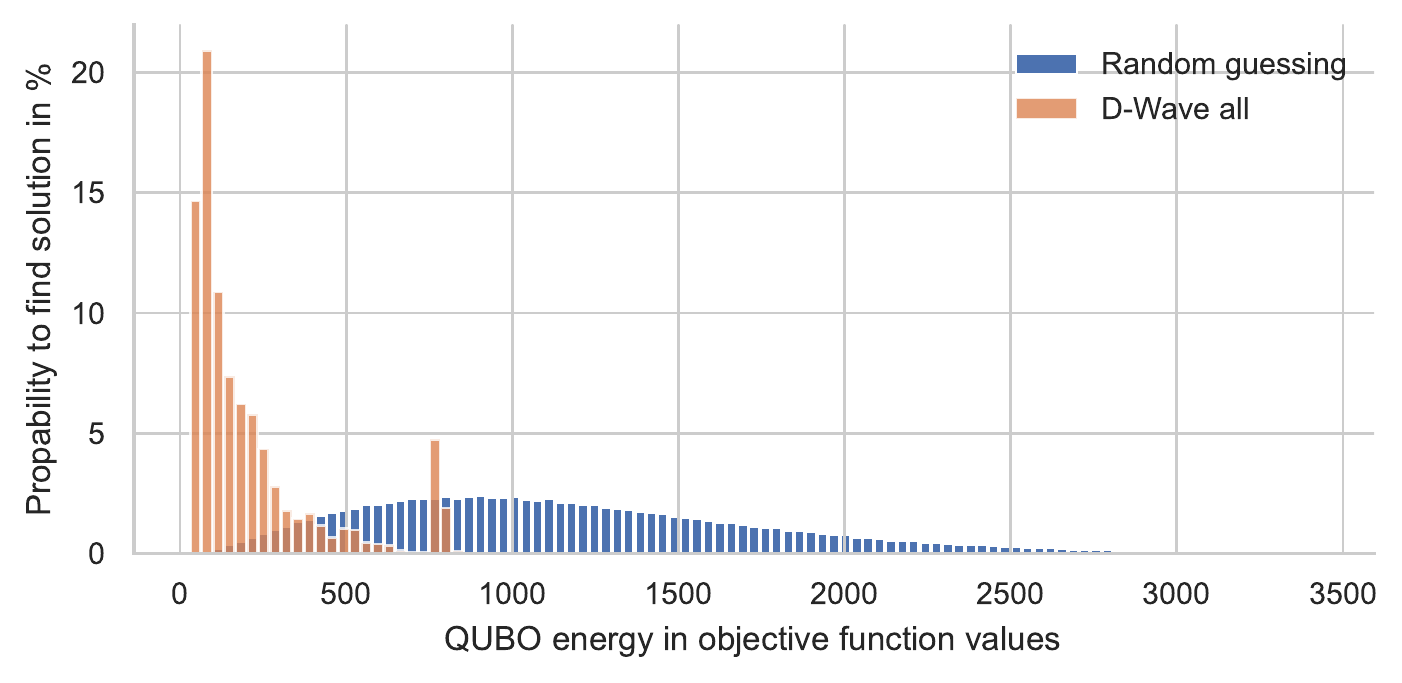}};
        \fill[white] (-4.4,-1.9)rectangle(-3.3,2.15); 
        \fill[white]
         (-3.5,-1.4)rectangle(4.5,-2.05); 
        \fill[white] (1.97,2.1)rectangle(4.4,-2.45); 
        \draw [gray!40,line width=0.45pt](-3.3,2)--(1.96,2);
        \node at (-3.275, -1.6)[anchor=base]{0};
        \node at (-2.2, -1.6)[anchor=base]{500};
        \node at (-1.175, -1.6)[anchor=base]{1000};
        \node at (-0.15, -1.6)[anchor=base]{1500};
        \node at (0.925, -1.6)[anchor=base]{2000};
        \node at (1.95, -1.6)[anchor=base]{2500};      
        
        \node at (-0.65, -1.8)[]{QUBO Energy in Objective Function Values};
        \node at (-3.4,1.7)[anchor=east, align=right]{20};
        \node at (-3.4,0.94)[anchor=east, align=right]{15};
        \node at (-3.4,0.18)[anchor=east, align=right]{10};
        \node at (-3.4,-0.59)[anchor=east, align=right]{5};
        \node at (-3.4,-1.35)[anchor=east, align=right]{0};
        \node at (-3.9,0.2)[align = left, rotate = 90]{Obtained Solution Density in \%};
        \draw [fill=white] (-1.1,1.1) rectangle (1.85,1.9);
        \node at (-1,1.52) [anchor=west,align= left]{\includegraphics[trim=102.9mm 57mm 33mm 5mm, clip,width=5mm]{energy.pdf}};
        \node at (-.35,1.65)[anchor=west, align=left]{Random Guessing};
        \node at (-0.35,1.32)[anchor=west, align=left]{D-Wave Advantage};
        
    \end{tikzpicture}
    \caption{Solution density with respect to \ac{QUBO} objective function values according to \eqref{eq:qubo_energy} for a three-node network. Density values correspond to all obtained D-Wave samples independent of the parameter setup and feasibility of the solution (orange) and to randomly sampled integers (blue) for $10^6$ samples. The density is defined as the number of samples within a bin divided by the number of all obtained samples; for each category respectively.}
    \label{fig:energy}
    \vspace{-2mm}
\end{figure}

The aforementioned evaluation procedure is a baseline to study the solvability and  dependencies on annealing parameters, numerical accuracy and various embeddings of the considered network problem for a three-node network.
Fig.~\ref{fig:energy} shows the energy distribution of the \ac{QUBO}'s objective function according to \eqref{eq:qubo_energy}.
It displays, on the one hand, solution vectors obtained by the annealing process for various parameter settings (orange).
On the other hand, it displays solutions obtained by a (uniform) random bit guessing approach (blue).
We observe that quantum annealing provides solution vectors with significantly lower energy values that follow a different distribution.

\begin{figure}
    \centering
    \begin{tikzpicture}[font=\sffamily\fontsize{7pt}{7pt}\selectfont, inner sep=0pt]
        \useasboundingbox (-3.95,-2.75) rectangle (4.9,2.9);
        \node at (0,0)[align=left, inner sep=0pt]{
            \includegraphics[width=\columnwidth]{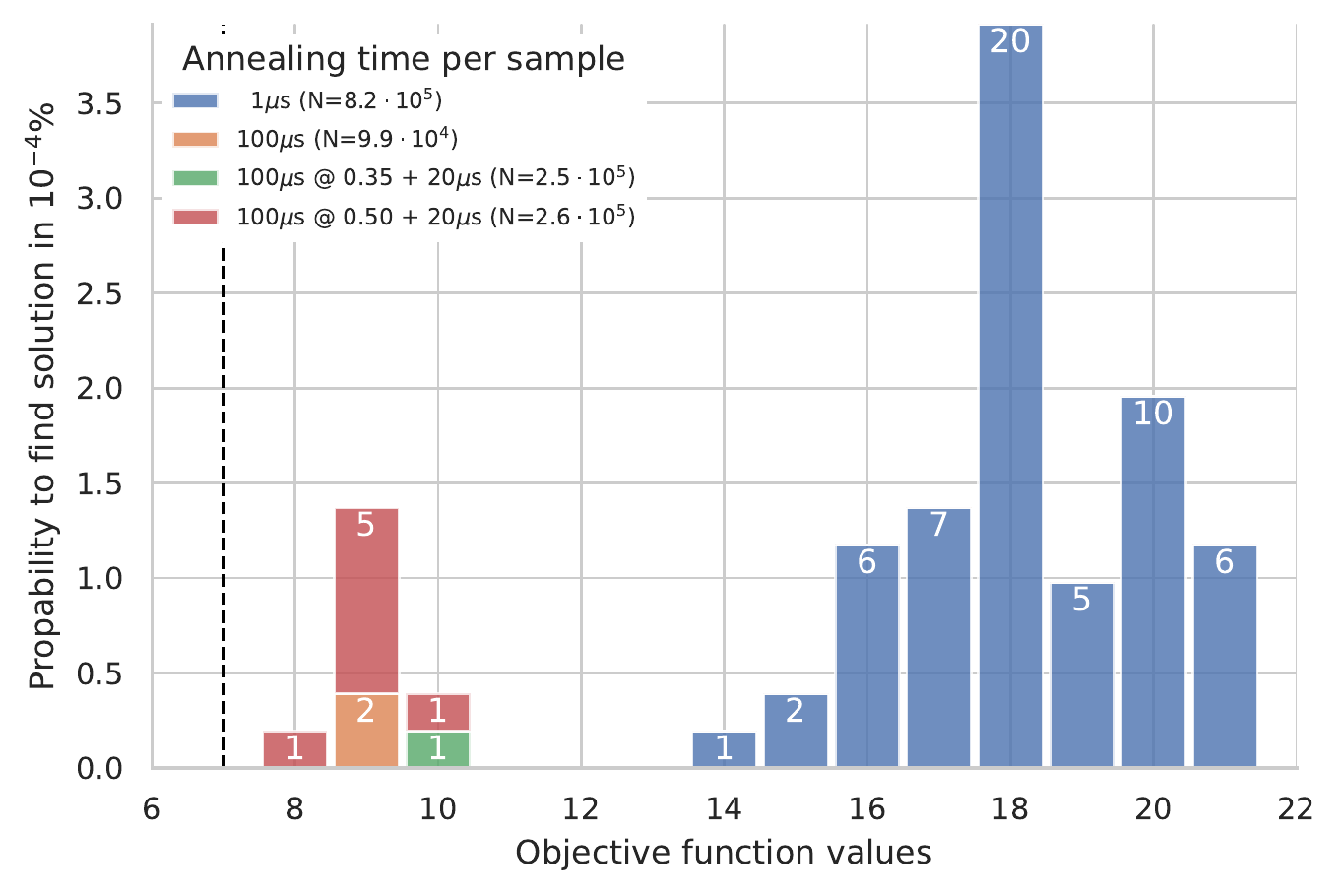}};
        \fill[white]
        (-4.3,-2.4)rectangle(-3.5,2.5); 
        \fill[white]
        (-3.7,-2.2)rectangle(4.3,-2.9); 
        \draw [gray!40,line width=0.5pt](-3.43,2.8)--(4.15,2.8);
        \node at (-3.45, -2.4)[anchor=base]{6};
        \node at (-2.975, -2.4)[anchor=base]{7};
        \node at (-2.5, -2.4)[anchor=base]{8};
        \node at (-2.025, -2.4)[anchor=base]{9};
        \node at (-1.55, -2.4)[anchor=base]{10};
        \node at (-1.075, -2.4)[anchor=base]{11};
        \node at (-0.6, -2.4)[anchor=base]{12};
        \node at (-0.125, -2.4)[anchor=base]{13};
        \node at (0.35, -2.4)[anchor=base]{14};
        \node at (0.825, -2.4)[anchor=base]{15};
        \node at (1.3, -2.4)[anchor=base]{16}; 
        \node at (1.775, -2.4)[anchor=base]{17};     
        \node at (2.25, -2.4)[anchor=base]{18};
        \node at (2.725, -2.4)[anchor=base]{19}; 
        \node at (3.2, -2.4)[anchor=base]{20}; 
        \node at (3.675, -2.4)[anchor=base]{21};
        \node at (4.15,-2.4)[anchor=base]{22}; 
        
        \node at (0.4, -2.6)[]{Objective Function Values};
        \node at (-3.5,1.65) [anchor=east, align=right]{3};
        \node at (-3.5,0.4) [anchor=east, align=right]{2};
        \node at (-3.5,-0.85) [anchor=east, align=right]{1};
        \node at (-3.5,-2.1)[anchor=east, align=right]{0};
        \node at (-3.8,0.3)[align = left, rotate = 90]{Feasible Solutions per Million Samples};
        \foreach \y in {-2.12,-0.9,0.32,1.55,2.78}{
            \draw[gray!40, line width= 0.8 pt] (4.13,\y)--(4.18,\y);}
        \node at (4.22,2.78) [anchor=west, align=left]{20};  
        \node at (4.22,1.55) [anchor=west, align=left]{15};
        \node at (4.22,0.32) [anchor=west, align=left]{10};
        \node at (4.22,-0.9) [anchor=west, align=left]{5};
        \node at (4.22,-2.12)[anchor=west, align=left]{0};
        \node at (4.7,0.3)[align = left, rotate = -90]{Feasible Solutions within all Samples};
        \draw [very thick, green!70!black!100] (-2.95,-2.1)-- (-2.95,2.8);
        \node at (-3.1,-0.65)[rotate=90, green!70!black!100]{Reference Value (CPLEX)};
        \draw [fill=white] (-3.36,0.8) rectangle (-0.15,2.73);
        \node at (-3.3,1.61) [anchor=west,align= left]{\includegraphics[trim=17.5mm 69mm 115.5mm 9.5mm, clip,width=5.3mm]{feasible-energy.pdf}};
        \node at (-2.7,2.2)[anchor=west, align=left]{1\,\textmu s};
        \node at (-2.7,1.8)[anchor=west, align=left]{100\,\textmu s};
        \node at (-2.7,1.4)[anchor=west, align=left]{100\,\textmu s @ 0.35 + 20\,\textmu s};
        \node at (-2.7,1.0)[anchor=west, align=left]{100\,\textmu s @ 0.50 + 20\,\textmu s};
        \node at (-3.25,2.5)[anchor=west, align=left]{Annealing Time/Schedule};
        
    \end{tikzpicture}
    \caption{%
        Solution density with respect to the objective function values \eqref{eq:cost_function} of the \ac{ILP} for a three-node network obtained by the quantum annealer.
        Solutions are classified as feasible solutions in post-processing by evaluating \eqref{eq:feasibility_check}.
        The density is defined as the number of feasible samples for an objective value divided by the number of all obtained samples over all submits (including parameter configurations which did not find feasible solutions).
        The samples are categorized by different annealing schedules, specified by the annealing time per sample (without overhead), and the number within the bars is the count for a given category.
        The green line is a known reference solution obtained by the classical ILP-solver CPLEX.}
    \label{fig:feasible-energy}
    \vspace{-2mm}
\end{figure}

A more detailed view is seen by analyzing the feasibility of solution samples, determined by calculating \eqref{eq:feasibility_check} post anneal, and evaluating the cost value $\ve c ^\top\ve x$.
Fig.~\ref{fig:feasible-energy} shows the amount (white numbers) of obtained feasible solutions per integer cost values (x-axis).
\emph{\textbf{The best solution we have found, corresponding to a cost value of 8, is one cost step apart from the best reference solution with a cost of 7.}}
The results can be further grouped according to the applied annealing times. 
Here we see that results within the category of a $100$ \textmu s annealing duration are better than for the $1$\textmu s case.
\emph{\textbf{Overall, feasible solutions are very rare, at the order of 13 per million if compared with all obtained samples.}} This obtained ratio determines the values of the y-axis of Fig.~\ref{fig:feasible-energy}.
At this point it is not possible for us to determine if the choice of the penalty term, the annealing schedule, other parameters, or their interplay are responsible for the quality and frequency of the optimal solution within a suitable region.
We are actively investigating this point.

\subsection{Sources of Errors and Statistics}

In theory, optimal solutions of the \ac{QUBO} problem are automatically optimal solutions to the network \ac{ILP} problem.
However, solutions returned by the \ac{QA} may be non-optimal because of errors during the annealing process.
For example, depending on the specification of the annealing schedule, non-adiabatic transitions from the ground state to an excited state may occur if the annealing happens too fast, and especially if the energy gaps between excited and ground states are small.

If the annealing happens too slowly, temporal or thermal decorrelation of qubits may occur so that qubits may freeze in position independent of the problem Hamiltonian.
On the other hand, it may not be possible to set up the problem on the hardware---even if a valid embedding was found.
For example, if qubit chains are too long, they potentially break such that one solves a different problem.
Or, since the magnetic fields generated by the hardware are only manipulated with finite precision, the resolution of the \ac{QUBO} entries may exceed the magnetic field precision.
It is \emph{a priori} not known if these problems may occur during an annealing schedule.
For a given problem setup, one must therefore perform multiple annealing schedules and post-process the results to identify valid solutions.
This procedure provides distributions of solutions with varying degrees of quality.

\section{Scalability Study}
\label{sec:scalability-study}
\subsection{Embedding of QUBO on the D-Wave Advantage\TM}
\label{sec:Embedding}
The process of determining a valid mapping of the \ac{QUBO} graph to the hardware graph is an essential constraint for being able to solve a problem.
The possibility of finding such a mapping, also called an embedding, is thus limited by the number of logical qubits present in the formulation, i.e., the dimension of the \ac{QUBO} matrix.
Furthermore, even if the number of available qubits is in principle sufficient, the density of the \ac{QUBO} matrix, defined as ratio of non-zero entries divided by its dimension squared, cannot exceed the total connectivity of the hardware graph.
Because the hardware topology may have a different connectivity as required by the \ac{QUBO}, several hardware qubits must be chained together to form a single logical qubit (nodes of the \ac{QUBO} graph) --- which places stronger constraints on the problem sizes.

This pre-processing procedure, e.g. finding a valid hardware embedding for the given problem, must be performed before submitting problems to the annealer.
Depending on the problem size, finding a valid embedding may take several hours on a single CPU.
However, since an embedding is a map from \ac{QUBO} nodes to hardware topology nodes, which does not depend on the relative coupling strengths, it is possible to export valid embeddings which are reusable for a larger set of problems.
Thus the expensive part of finding an embedding must be performed only once for a class of \ac{QUBO} problems.
Generally, embeddings which have small chains while using the entire hardware are found to be more optimal\cite{bib:Willsch2021ajl}.

\subsection{Scaling of Network Sizes}

\begin{figure}
    \centering
    \includegraphics[width=0.4\columnwidth]{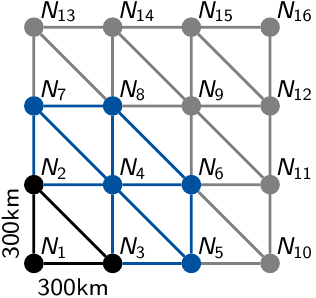}
    \vspace{-2mm}
    \caption{Topology of a growing network. (\protect\colorlabelcircle{black}): smallest network with 3 nodes, (\protect\colorlabelcircle{uni-dark}): exemplary extension to a 8 node network, (\protect\colorlabelcircle{gray}) maximal considered extension.}
    \label{fig:growing_network}
    \vspace{-4mm}
\end{figure}

For studying the scalability in terms of the communication network size, a growing network topology is used as depicted in Fig.~\ref{fig:growing_network}.
The nodes of this topology are added in ascending order starting with a three-node network ($N_1,N_2$ and $N_3$) in the lower left corner. 
The edges are added according to the shown topology, such that added nodes are connected with the already existing network. 
Extending to a four-node network $N_4$ adds the edges $N_2$--$N_4$ and $N_3$--$N_4$, for example.
The vertical and horizontal distance between neighboring nodes is set to 300~km.

Finally, a set of topologies with $|V| \in \{3,4, \ldots, 16\}$ nodes was studied to test the feasibility of finding an embedding. 
The network loads are defined by a demand matrix with values $h_d$ for traffic demands between the pairs of disjunct nodes $d\in D$. 
The values for $h_d$ are sampled from a normal distribution $\mathcal{N}(\mu,\sigma)$ with $\mu=75$ and $\sigma=20$ Gbit/s. 
We used the values in Table~\ref{tab:param_scaling} for the parametrization of our ILP for the QUBO mapping procedure.

\begin{table}
    \caption{Selected Parameters for Various Considered Network Sizes.}
    \label{tab:param_scaling}
    \centering
    \renewcommand{\tabcolsep}{4pt}
    \begin{tabular}{lcccccc}
        \toprule
        Network Size (\# Nodes)         & $|V|$                 & 3--4 & 5--7 & 8--11 & 12 & 13--16 \\
        \midrule
        Installed Transceivers per Node & $\eta_\mathrm{max}$   & 15   & 31   & 63    & 63 & 127    \\
        Max.\ Capacity of Circuit Paths & $\omega_\mathrm{max}$ & 3    & 3    & 3     & 7  & 7      \\
        \bottomrule
    \end{tabular}
\end{table}

\subsection{Hardware Limitations of the used Quantum Annealer}
\begin{figure}
    \centering
    \begin{tikzpicture}[font=\sffamily\fontsize{7pt}{7pt}\selectfont, inner sep=0pt]
     \useasboundingbox (-3.7,-2.4) rectangle (4.6,2.2);
        \node at (0,0)[align=left, inner sep=0pt]{
    \includegraphics[width=1\columnwidth]{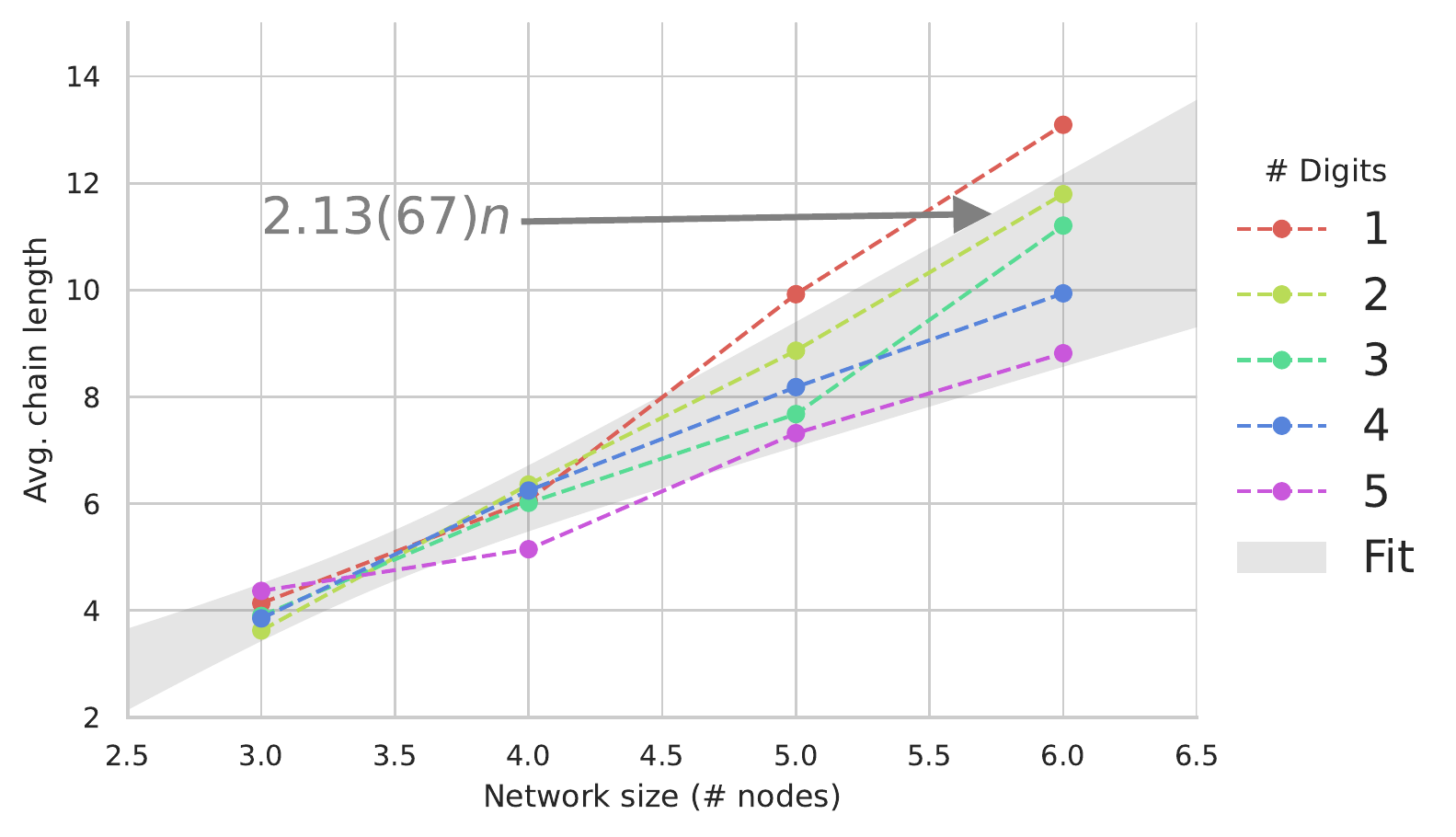}};

\fill[white] (-4.4,-1.9)rectangle(-2.86,2.15); 
\fill[white] (-4,-1.9)rectangle(4,-2.45); 
\fill[white] (-4,2.1)rectangle(4,2.45); 
\fill[white] (2.05,2.1)rectangle(4.4,-2.45); 

\node at (-2.85, -2.1)[anchor=base]{3};
\node at (-1.22, -2.1)[anchor=base]{4};
\node at (0.42, -2.1)[anchor=base]{5};
\node at (2.05, -2.1)[anchor=base]{6};

\node at (-0.36, -2.3)[]{Network Size, Number of Nodes \textit{$|$V$|$}};

\fill[white] (-2.83,1) rectangle(-2.05,1.4);
\fill[white] (-2.02,1) rectangle(-1.25,1.4);
\node at (-1.8,1.2)[]{2.13$\cdot|$\textit{V}$|$};

%
\node at (-2.9,2.1)[anchor=east, align=right]{14};
\node at (-2.9,1.45)[anchor=east, align=right]{12};
\node at (-2.9,0.8)[anchor=east, align=right]{10};
\node at (-2.9,0.15)[anchor=east, align=right]{8};
\node at (-2.9,-0.5)[anchor=east, align=right]{6};
\node at (-2.9,-1.15)[anchor=east, align=right]{4};
\node at (-2.9,-1.8)[anchor=east, align=right]{2};
\node at (-3.5,0)[align = left, rotate = 90]{Average Chain Length};
\node at (2.3,0) [anchor=west,align= left]{\includegraphics[trim=135mm 28mm 12mm 22mm, clip,width=0.76cm]{embedding-scaling.pdf}};
\node at (3.1,1.0)[anchor=west, align=left]{1};
\node at (3.1,0.6)[anchor=west, align=left]{2};
\node at (3.1,0.2)[anchor=west, align=left]{3};
\node at (3.1,-.2)[anchor=west, align=left]{4};
\node at (3.1,-.6)[anchor=west, align=left]{5};
\node at (3.1,-1)[anchor=west, align=left]{Linear Fit};
\node at (2.3,1.5)[anchor=west, align=left]{Numeric Accuracy \textit{a},\\i.e.\ Number of Digits};

    \end{tikzpicture}
    \caption{Average chain length (physical qubits needed to form a logical qubit) depending on the number of nodes in the network and digits required to represent demands. The gray band represents a one-parameter fit with propagated uncertainties.}
    \label{fig:scaling}
\end{figure}

\enlargethispage{4mm}
The optimization problem size, i.e. the dimension of the \ac{QUBO}, can be expressed by the number of required logical qubits. As logical qubits are realized by a coupled chain of physical qubits, the average chain length can be seen as a metric for the efficiency of an embedding. Fig.~\ref{fig:scaling} shows this metric for 3 to 6 node networks. The varied number of digits $a$ changes the numerical accuracy of the \ac{ILP}'s constraint \eqref{eq:const_circuits}, and changes therewith the problem size. We observe that the average chain length follows a linear law of $2.13\cdot|V|$.
Using the default embedding algorithm (minorminer), we were only able to find embeddings for networks with six or less nodes for the D-Wave Advantage\TM 5.2/5.3.
Assuming the observed chain-length scaling is valid for future \ac{QA} hardware, it can be used to extrapolate to larger network sizes with more than six nodes. 

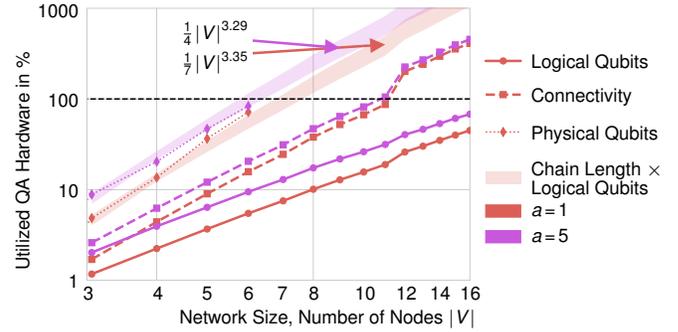
\begin{figure}
    \centering
    \begin{tikzpicture}[font=\sffamily\fontsize{7pt}{7pt}\selectfont, inner sep=0pt]
        \useasboundingbox (-4.2,-2.2) rectangle (4.45,2.2);
        
        \node at (0,0)[]{\includegraphics[width=1\columnwidth]{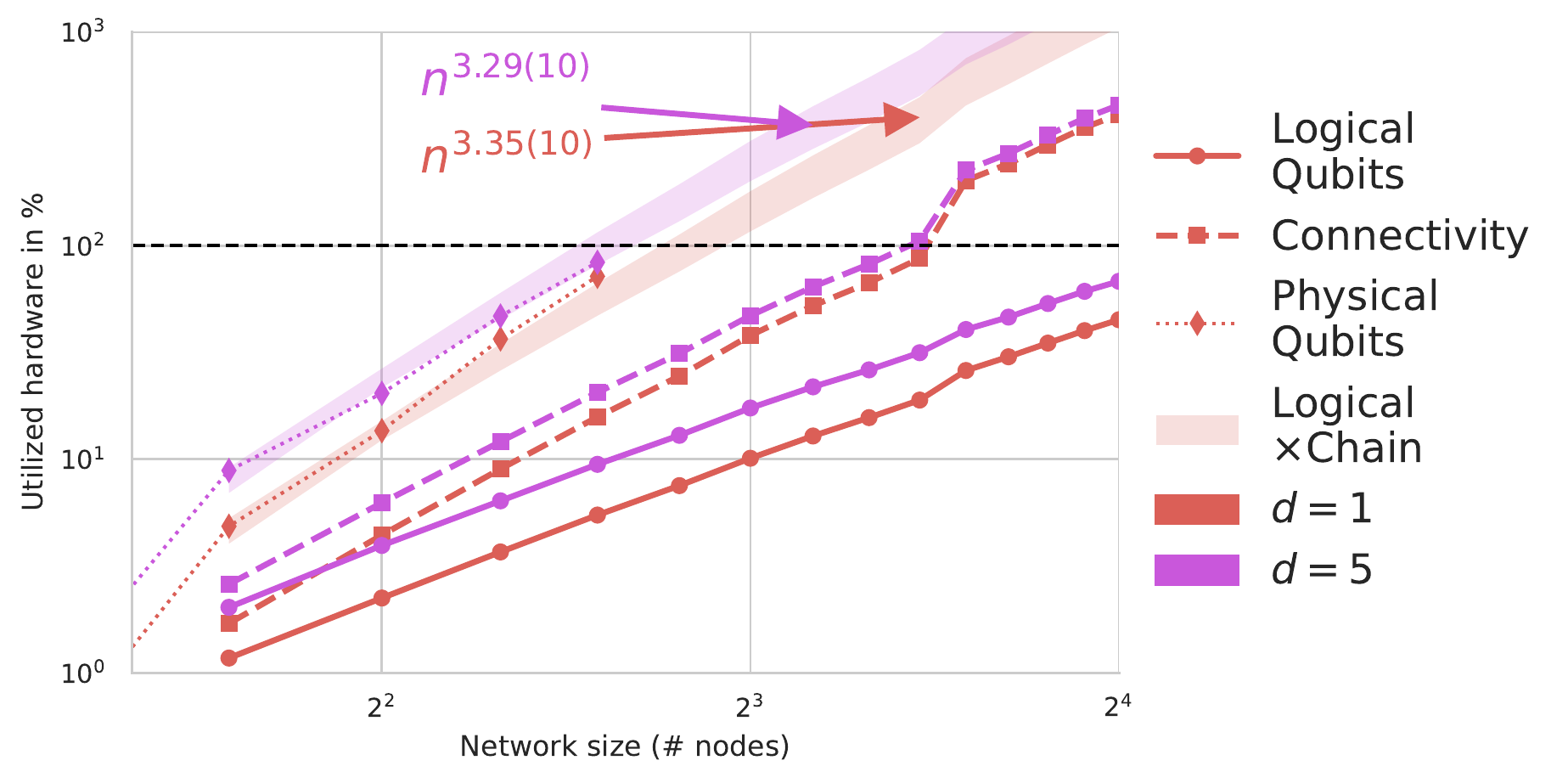}};
        
        \fill[white] (2.7,-1.5)rectangle(4.3,2);
        \fill[white] (-4,-1.6)rectangle(4,-2.15);
        \fill[white] (-4.4,-1.8)rectangle(-3.2,2.15);
        \node at (2.7,1.3)[anchor=west, align = left, fill=white]{Logical Qubits};
        \node at (2.7,0.85)[anchor=west, align = left, fill=white]{Connectivity};
        \node at (2.7,0.35)[anchor=west, align = left, fill=white]{Physical Qubits};
        \node at (2.7,-0.25)[anchor=west, align = left, fill=white]{Chain Length $\times$\\Logical Qubits};

        \node at (2.7,-0.65)[anchor=west, align = left, fill=white]{\textit{a}\,=\,1};
        \node at (2.7,-1)[anchor=west, align = left, fill=white]{\textit{a}\,=\,5};
        \fill [white] (-2.2,1)rectangle(-1,2);
        \node at (-1.5,1.7)[]{$\frac{\text{1}}{\text{4}}|$\textit{V}$|^\text{3.29}$};
        \node at (-1.5,1.3)[]{$\frac{\text{1}}{\text{7}}|$\textit{V}$|^\text{3.35}$};
        
        \node at (-3.18, -1.85)[anchor=base]{3};
        \node at (-2.25, -1.85)[anchor=base]{4};
        \node at (-1.6, -1.85)[anchor=base]{5};
        \node at (-1.05, -1.85)[anchor=base]{6};
        \node at (-0.6, -1.85)[anchor=base]{7};
        \node at (-0.18, -1.85)[anchor=base]{8};
        \node at (0.5, -1.85)[anchor=base]{10};
        \node at (1.05, -1.85)[anchor=base]{12};
        \node at (1.5, -1.85)[anchor=base]{14};
        \node at (1.9, -1.85)[anchor=base]{16};
        \node at (0, -2.1)[]{Network Size, Number of Nodes $|$\textit{V}$|$};
        
        \draw [gray!40,line width=0.45pt](-3.2,-1.59)--(-3.2,2.03);
        \node at (-3.3,2)[anchor=east, align=right]{1000};
        \node at (-3.3,0.8)[anchor=east, align=right]{100};
        \node at (-3.3,-0.4)[anchor=east, align=right]{10};
        \node at (-3.3,-1.6)[anchor=east, align=right]{1};
        \node at (-4.05,0)[align = left, rotate = 90]{Utilized QA Hardware in \%};
    \end{tikzpicture}\vspace{-2mm}
    \caption{Hardware utilization of problem depending on problem size. The hardware utilization is expressed by the logical qubits (filled circles) corresponding to the dimension of the \ac{QUBO} (at most 5.6k qubits for a perfect embedding), the connectivity of the \ac{QUBO} (squares) corresponding to non-zero entries in the \ac{QUBO} (at most 40.1k), and physical qubits (diamonds) corresponding to logical qubits times the average chain length (at most 5.6k qubits). Results are displayed for the best scaling using a demand float precision of one bit (red) and the most precise scaling with five bit precision (purple).}
    \label{fig:hardware}
    \vspace{-4mm}
\end{figure}

Fig.~\ref{fig:hardware} shows the resulting relative hardware utilization of the quantum annealer for network problems with 3 to 16 nodes. The D-Wave Advantage 5.2\TM provides in total 5600 physical qubits and roughly 40100 coupling elements. This marks the 100\% line of hardware utilization. The amount of required logical qubits increases with the network size in a range from 66 to 3822, i.e.\ 1.1\% to 66.4\%.
Further, the \acp{QUBO} of the network problems require qubit connecting couplers on the logical level in the range of 684 to 182110, i.e. these are the non-zero elements of $\ve{Q}$. The amount of maximal available couplers is already exceeded for a network of 12 nodes.
As chains of physical qubits must be built by coupling to realize a single logical qubit, the amount of physical available qubits and/or coupling element is already exceeded for networks with 7 or more nodes. The required amount of physical qubits can be obtained by multiplication of the average chain length and amount of logical qubits, e.g. the 6 node network with a 5 digit accuracy $a$ has $532\cdot 8.8=4682$ physical qubits.

We summarize that the procedure of finding an embedding for network problems on the D-Wave's quantum annealer forms a stronger constraint than the constraints imposed by the naive estimations related to the \ac{QUBO} size and density.
Specifically, for the largest problem we were able to embed, the amount of physical qubits needed was roughly an order of magnitude larger than the naive estimate based on the size of the \ac{QUBO}.
Increasing the problem accuracy increases the problem size and therefore the required \ac{QPU} hardware. This effect is not as strong as adding further nodes to a network. Thus, \emph{\textbf{the largest impact on the problem complexity comes from the number of network nodes.}} Extrapolation of the physical qubit curve in Fig.~\ref{fig:hardware} allows us to estimate how large a \ac{QPU} in terms of physical qubits for real network problems should be.  We find that \emph{\textbf{at least 10 times more physical qubits are required to operate a wide-area network with a reliable size of 12 to 16 nodes by quantum annealing.}}
An adjustment of the topology, i.e., an increase of individual qubit's connectivity would also enable access to larger networks, which is however more difficult to quantify.

\section{Conclusion and Outlook}
\label{sec:conclusion}
\enlargethispage{1\baselineskip}

We proposed an algorithmic approach relying on the cutting-edge technology of quantum computing for the optimization of resource utilization in \ac{SDN}-controlled optical wide-area networks.
The optimization problem is modeled by an \ac{ILP}, translated in several steps to a \ac{QUBO}, which can be solved on a quantum annealer like the D-Wave Advantage\TM 5.2/5.3.
The algorithm can be used for traffic engineering, resource allocation and restoration.

We studied the quantum-based solvability of a three node network. Compared to a random guessing method, solutions obtained by the quantum annealer show a significant lower energy on average.
Solution vectors are checked for feasibility and compared with a reference solution obtained by CPLEX (classical \ac{ILP}-solver).
We showed that feasible solutions with cost values close (one unit-step less optimal) to the reference solution are obtainable.
At the current stage, feasible reconfigurations close to the optimal value for a three-node problem can be obtained every minute. This number is defined by the number of samples needed to find a feasible solution and the total run time per sample for an individual anneal.
For a given effective anneal time $t_\mathrm{eff} \sim t_\mathrm{ps}+t_\mathrm{p} [\text{\textmu s}]$, we estimate the total run time per sample by $0.58+5.75\cdot t_\mathrm{eff} [\text{ms}]$, which is based on experience values. In case of $t_\mathrm{eff}\sim120$ \textmu s, we require $N\sim 5\cdot10^4$ samples with a total run time of 1.27 ms each, such that a feasible solution can be found in 63.5 s. 

Further, we discovered that a setting with penalty of 4 and annealing time of $100$\,\textmu s achieved the best results.
We see indications that some annealing parameter configurations have a higher chance to return feasible solutions---potentially allowing one to find feasible solutions within less than a minute of run time.
While decreasing the length of an annealing schedule to a few \textmu s in principle allows one to obtain solutions more frequently, at some point the problem setup time dominates the effective annealing time.
As such, there will be an optimal value of annealing parameters which minimize the time to solution.
The parameter variations for the annealing process has to be studied further as the current solution set represents only an empirical sample set which does not allow one to formulate strong statement on statistical correlations.

Our feasibility study shows that the proposed network problem for up to 6 nodes can be embedded on the D-Wave \ac{QPU}. We estimate that the amount of physical qubits, assuming the same qubit connectivity, should be in the range of 50000 or above to optimize networks in reasonable sizes (12 to 16 nodes).
Finally, as larger Ising model based solvers with up to 100000 qubits \cite{bib:100kqubit} are in the reach, embedding 12 to 16 node networks seems realistic in the near future.

For realistic networks, classical heuristics find solutions within several minutes.
Compared to classical heuristics, the length of individual annealing schedules are independent of the problem size. 
However, the probability of finding a feasible and optimal solution will be smaller than the probability we have obtained for the three-node problem due to the increased sample space.
Thus, the focus of future works should concentrate on both, finding more optimal ways to embed problems to scale to larger networks and finding optimal setup parameters to decrease the time to solution.

As quantum computing can be superior against classical computing, demonstrated in \cite{bib:quantum_supremacy}, a further speed-up might be achievable by newer generations of quantum computers or a problem specific \ac{QPU}, such that network reconfiguration within seconds or below might be possible.\enlargethispage{1\baselineskip}
It remains to be shown that obtaining feasible solutions within a reasonable time is possible for such larger networks.
If successful, the proposed approach for network optimization has the potential to have a large impact on how networks are operated in the future and ultimately enable real-time network automation.

\appendices
\section{Basic Principles behind the used D-Wave Machine}
\label{sec:app_dwave}

Here we provide a short discussion on the D-Wave architecture and the basic principles underlying the quantum annealing process. For further details we refer to \cite{bib:DWAVE_Annealing}.

\subsection{The Ising Model}
\begin{figure}
    \includegraphics[width=\columnwidth]{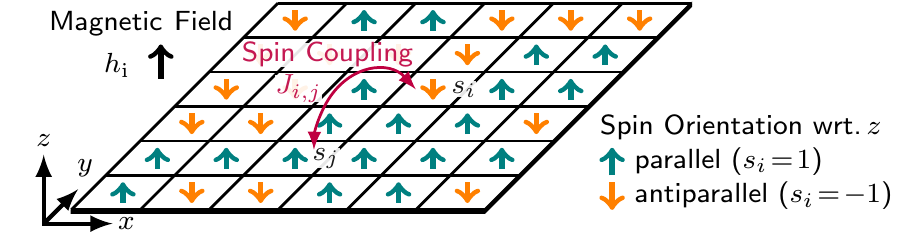}
    \caption{Visualization of spin configuration in a 2D-lattice (flat material) which is determined by the Ising model \eqref{eq:hamiltonian_ising}.}
    \label{fig:ising}
\end{figure}

Central to the annealing process is the ability to manipulate a 2-D array of 2-state spins that form the so-called Ising model, as depicted in Fig.~\ref{fig:ising}.

In physics, the Hamiltonian function that describes this Ising system is given by
\begin{equation}
    \label{eq:hamiltonian_ising}
    \mathcal{H}_\mathrm{Ising}(\ve s) =
    \sum\limits_i h_i s_i+
    \sum\limits_{i>j} J_{i,j}s_{i}s_{j}\ ,
\end{equation}
where $s_i=\pm1$ are spin projections in the $z$ direction and $h_i$ is an external magnetic field at site $i$. The coupling between spins at sites $i$ and $j$ is given by $J_{i,j}$.
This expression represents the total energy of the system and can be used to derive its equations of motions .
As the model represents an array of spins in 2-D, it provides a simple representation of ferromagnetism that exhibits a second order phase transition.   From a QA perspective, it serves as an objective function to be minimized.

\subsection{Quantum Annealing as Hardware Process}

The quantum mechanical equivalent of a classical system can be obtained by replacing canonical coordinates and momenta with operators (canonical quantization).
As a consequence, previously commuting expressions may now not be commutable.
In case of the Ising model, the spin variables $s_i$ are replaced by spin operators: the Pauli matrices $\hat \sigma$.
The D-Wave Advantage\texttrademark\ system starts initially with a \emph{transverse} Ising model.
By the application of external magnetic fields, the annealer adjusts the the time-dependent amplitudes of the Hamiltonian operator in both the $x$, or transverse, direction and $z$ direction, respectively,

\begin{equation}
    \hat{\mathcal{H}}_\mathrm{QPU} =    \frac {A(s)} {2}
    \hat{\mathcal{H}}_\mathrm{initial} +
    \frac{B(s)}{2}\hat{\mathcal{H}}_\mathrm{problem}
    \label{eq:hamiltonian_annealing}
\end{equation}
\begin{equation}
    \hat{\mathcal{H}}_\mathrm{initial} = \sum_i\pauli{x}{i}
\end{equation}
\begin{equation}
    \hat{\mathcal{H}}_\mathrm{problem} = \sum\limits_i h_i \pauli{z}{i} +
    \sum\limits_{i>j} J_{i,j}\pauli{z}{i}\pauli{z}{j}\ .
\end{equation}

\begin{figure}
    \includegraphics[width=\columnwidth]{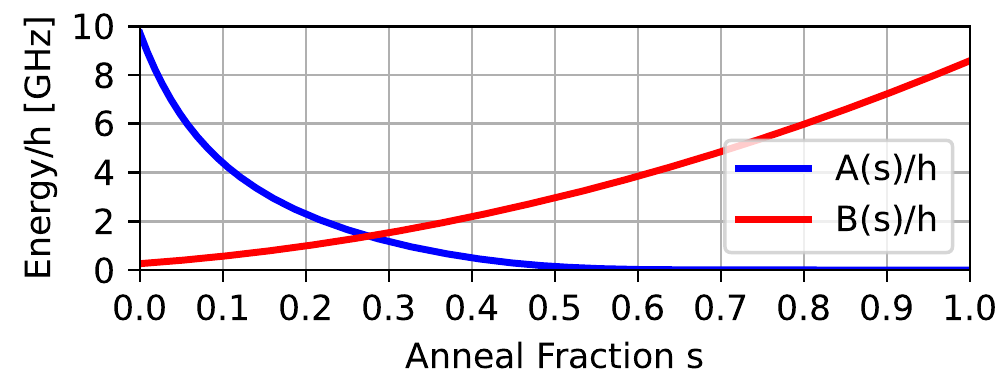}
    \caption{Energy variation of $A(\rm s)$ and $B(\rm s)$ to perform the annealing within the D-Wave Advantage\TM 5.2 acc to \eqref{eq:hamiltonian_annealing}. Energy values are normalized by the Planck constant $h=6.62\cdot10^{-34} \mathrm{Joules}$ based on the relation $E=hf$. The anneal fraction $s$ will be varied over time within the annealing duration to perform either a standard, paused, quenched, reverse or individual annealing schedule.}
    \label{fig:energy_annealing}
\end{figure}

Here $\sigma_{x,z}$ represent Pauli sigma matrices.  Because the spin projections in the $x$ and $z$ directions do not commute, this system must be solved quantum mechanically and is therefore also known as the \emph{quantum} Ising model.   The external fields depend on time.  Initially $A(\rm s)\gg$ $B(\rm s)$ at time $t=0$ corresponding to an annealing fraction $\rm{s}=0$, but adiabatically changes to $B(\rm s)\gg$ $A(\rm s)$ at some anneal time $t=t_f$ corresponding to an annealing fraction $\rm s=1$, as shown in Fig.~\ref{fig:energy_annealing}.  After time $t_f$ the system has annealed to purely $\hat{\mathcal{H}}_\mathrm{problem}$, and since $\hat\sigma_z$ has either $\pm1$ expectation values, this is identical to the classical Ising model in \eqref{eq:hamiltonian_ising}.

\subsection{The QUBO and Ising Model as Interface to Program the Problem to be Solved on the D-Wave Quantum Annealer}

The programming of D-Wave’s quantum annealer requires the representation of the problem as a \ac{QUBO} problem for bit vectors $\ve{q} \in \{0, 1\}^N$
\begin{equation}
    X^2(\ve{q})=\ve{q}^\top \ve{Q} \ve{q} \,.
\end{equation}
The annealer aims at finding the optimal bit vector $\ve{q}$ which minimizes the objective function $X^2$.

The Ising model wave function components $\sigma_i$ are related to components of the \ac{QUBO} bit vectors $\ve q$ by a linear shift $\sigma_i = 2 q_i - 1$.
Equating respective objective functions allows to identify the mapping from \ac{QUBO} matrix to the Ising Hamiltonian in matrix form
\begin{equation}
    \ve{q}^\top \ve{Q} \ve{q}
    =
    \ve{\sigma}^\top \ve{J} \ve{\sigma}
    +
    \ve{h}^\top \ve{\sigma}
    +
    g\, ,
\end{equation}
which is valid for
\begin{equation}
    \begin{aligned}
        \ve{J} & = \frac{1}{4} \ve Q_\mathrm{TL}                                                               \\
        \ve{h} & = \frac{1}{2} \ve q_\mathrm{T} + \frac{1}{2}  \ve Q_\mathrm{TL} \ve{1}                        \\
        g      & = \frac{1}{4}\ve{1}^\top \ve Q_\mathrm{TL} \ve 1 + \frac{1}{2} \ve 1^\top \ve q_\mathrm{T}\,. \\
    \end{aligned}
\end{equation}
These equations were obtained by expressing \ac{QUBO} vectors $\ve q$ as Ising vectors $\ve \sigma$, defining the traceless part of the \ac{QUBO} $\ve Q_\mathrm{TL} = \ve  Q - \diag\{\ve q_\mathrm{T}\}$ with the trace vector $\ve q_\mathrm{T} = \diag^{-1}\{\ve Q\}$ of $\ve Q$ (since $\ve J$ is not allowed to have diagonal components), and using that $q_i^2 = q_i$.
Here $\diag\{\cdot\}$ transforms a vector into a diagonal matrix and $\diag^{-1}\{\cdot\}$ extracts the diagonal part of the matrix as a vector.
$\ve 1$ is a vector where each component is one.

Typically, the \ac{QUBO} matrix $\ve{Q}$ is symmetric ($\ve{Q}^\top=\ve{Q}$)  and is used only in the form $\ve{q}^\top \ve{Q}\ve{q}$ to determine a energy value. This allows one to transform $\ve{Q}$ to a triangular matrix $\ve Q_\mathrm{Triang}=\tril\{\ve Q\}^\top+\triu\{ \ve Q\}$, such that the problem can be also given in a reduced form according to
\begin{equation}
    X^2(\ve q)=\ve q^\top \ve Q_\mathrm{Triang} \ve q.
\end{equation}  Function $\tril\{\cdot\}$ selects the triangular part below the main diagonal of a matrix, and $\triu\{\cdot\}$ the upper triangular part inclusive the main diagonal.

\section{Derivation of the ILP to QUBO Problem Mapping}
\label{sec:app_problem_mapping}
For a given \ac{ILP} with introduced slack variables $\ve s$, which refactors inequality constraints as equality constraints,
\begin{equation}
    \begin{aligned}
         & \ve c^\top \ve x\rightarrow \min \\
         & \ve A\ve x+\ve b+\ve s=\ve 0\ ,
    \end{aligned}
\end{equation}
we can get rid of the constraints by further introducing a sufficiently large penalty factor $p\geq 1$. Thus, the objective function and constraints of the \ac{ILP} can be combined to a quadratic optimization of objective and penalty
\begin{equation}
    X^2(\ve{\x}, \ve{s})=\ve{c}^\top \ve{x} + p \norm{\mat{A}\ve{x}+\ve{b}+\ve{s}}^2
    \rightarrow \min \,.
\end{equation}
Given that the penalty term $p$ is large enough, the combined minimization of $X^2$ in terms of $\ve{\x}, \ve{s}$ returns the optimal vector $\ve{x}_0$ which minimizes $X^2$ under the constraints.

Integer variables in $\ve x$ and $\ve s$ have to be expressed in binary form as $\ve q _x$ and $\ve q _s$ 
according to 
\begin{equation}
    \ve{x}=\mat{Z}_x\ve{q}_x, \quad \ve{s}=\mat{Z}_s\ve{q}_s\,,
\end{equation}
see also \eqref{eq:z_indv}, \eqref{eq:z_comp} and \eqref{eq:q_vector}. Thus, we define the \ac{QUBO} as
\begin{equation}
    \hspace{-1.5mm}X^2(\ve{q})= \ve{c}^\top \mat{Z}_x\ve{q}_x + p \norm{\mat{A}\mat{Z}_x\ve{q}_x+\ve{b}+\mat{Z}_s\ve{q}_s}^2
    \rightarrow \min\,.
\end{equation}
By applying the rule $\norm{\epsilon}^2=\epsilon^\top \epsilon$, we obtain
\begin{equation}
\begin{aligned}
    &X^2(\ve{q})
    =\ve{c}^\top\ve Z_x\ve{q}_x 
    + p \ve{q}_x^\top \norm{\ve A\ve Z_x}^2\ve{q}_x 
    + p \ve{q}_x^\top\ve Z_x^\top \ve A^\top \ve{b}
    \nonumber\\
    &
    + p \ve{b}^\top \ve A \ve Z_x\ve{q}_x
    + p\ve{q}_s^\top \norm{\ve Z_s}^2\ve{q}_s
    + p\ve{q}_s^\top\ve Z_s^\top \ve{b}+ p\norm{\ve{b}}^2
    \\
    &
    + p\ve{b}^\top \ve Z_s\ve{q}_s 
    + p\ve{q}_s^\top\ve Z_s^\top \ve A\ve Z_x\ve{q}_x
    + p \ve{q}_x^\top\ve Z_x^\top \ve A^\top \ve Z_s \ve{q}_s\,.
\end{aligned}
\end{equation}
Some expressions like $\ve q_x^\top \ve Z_x^\top \ve A^\top \ve b$ can be equivalently expressed by $\ve q_x^\top \diag\{ \ve Z_x^\top \ve A^\top \ve{b} \}\ve q_x$ as $\ve Z_x^\top \ve A^\top \ve b$ is a vector and the binary vectors $\ve q_x$ and $\ve q_s$ will only selectively combine elements of the diagonal matrix during multiplication.
\begin{equation}
    \begin{aligned}   
    &X^2(\ve q) =
    p \ve q_s^\top \ve Z_s^\top \ve A\ve Z_x\ve q_x +
    p \ve q_x^\top \ve Z_x^\top \ve A^\top \ve Z_s \ve q_s\\
    &+p\ve q_x^\top \left[\frac{1}{p}\diag\{\ve{c}^\top \ve Z_x\}+ \diag\{ 2\ve b^\top \ve A \ve Z_x \}+ \norm{\ve A \ve Z_x}^2 \right]\ve q_x
     \nonumber\\
    &
    + p \ve q_s^\top \left[\norm{\ve Z_s}^2 + \diag \{ 2 \ve Z_s^\top \ve b\}\right]\ve q_s
    + p\norm{\ve b}^2\nonumber\\
    \end{aligned}
\end{equation}
Further refraction of the  $\ve q_x$- and $\ve q_s$-dependent terms and their mapping to the following  matrix form of the QUBO,
\begin{equation}
    X^2(\ve{q})=\ve{q}^\top \ve{Q} \ve{q}+ C = p\begin{bmatrix}
        \ve q_x\\
        \ve q_s
    \end{bmatrix}^\top 
    \begin{bmatrix}
        \ve Q_{xx} & \ve Q_{xs}\\
        \ve Q_{sx} & \ve Q_{ss}\\
    \end{bmatrix}
\begin{bmatrix}
    \ve q_x\\
    \ve q_s\\
\end{bmatrix} + C\nonumber\,,
\end{equation}
leads to the ILP formulation as QUBO problem given in \eqref{eq:qubo_matrix}.

\clearpage
\end{document}